\newif\ifjournal  
\newif\ifdraft    

\ifjournal
  \ifdraft
    \documentclass[acmsmall,screen,review]{acmart}
  \else
    \documentclass[acmsmall,screen]{acmart}
  \fi

  \AtBeginDocument{%
    }

  \acmDOI{10.1145/3772288}

  \setcopyright{cc}
  \setcctype{by}

  \acmJournal{TOMS}
  \acmYear{2025}
  \acmVolume{1}
  \acmNumber{1}
  \acmArticle{1}
  \acmMonth{1}
  \acmPrice{}
\else
  \documentclass[times]{article}

  \usepackage{geometry}
  \usepackage[square]{natbib}
  \usepackage{url}
  \setcitestyle{authoryear, square, yysep={;}}
  \setcitestyle{aysep={}}

  \usepackage{orcidlink}

  \usepackage{booktabs}
\fi

\usepackage{cleveref}
\usepackage[cachedir=minted-cache,frozencache]{minted2}
\setminted[cpp]{fontsize=\footnotesize,frame=single}
\usepackage{subcaption}
\usepackage{xcolor}
\usepackage{xspace}

\newcommand{\code}[1]{\texttt{#1}}
\newcommand{\arborx}{\textsc{ArborX}\xspace}

\definecolor{RedOrange}{HTML}{FF4433}
\definecolor{Cerulean}{HTML}{007BA7}
\definecolor{Plum}{HTML}{DDA0DD}
\definecolor{OliveGreen}{HTML}{808000}

\begin{document}

\makeatletter
\def\blfootnote{\xdef\@thefnmark{}\@footnotetext}
\makeatother

\title{The ArborX library: version 2.0}

\ifjournal
  \author{Andrey Prokopenko}
  \orcid{0000-0003-3616-5504}
  \email{prokopenkoav@ornl.gov}
  \author{Daniel Arndt}
  \orcid{0000-0001-8773-4901}
  \email{arndtd@ornl.gov}
  \author{Damien Lebrun-Grandi\'e}
  \orcid{0000-0003-1952-7219}
  \email{lebrungrandt@ornl.gov}
  \author{Bruno Turcksin}
  \orcid{0000-0001-5954-6313}
  \email{turcksinbr@ornl.gov}
  \affiliation{%
    \institution{Oak Ridge National Laboratory}
    \streetaddress{1 Bethel Valley Rd}
    \city{Oak Ridge}
    \state{Tennessee}
    \country{USA}
    \postcode{37830}
  }

  \renewcommand{\shortauthors}{Prokopenko et al.}



\else
  \author{
    A.~Prokopenko\thanks{Oak Ridge National Laboratory}\enskip\orcidlink{0000-0003-3616-5504},
    D.~Arndt\footnotemark[1]\enskip\orcidlink{0000-0001-8773-4901},
    D.~Lebrun-Grandi\'e\footnotemark[1]\enskip\orcidlink{0000-0003-1952-7219},
    B.~Turcksin\footnotemark[1]\enskip\orcidlink{0000-0003-0103-888X}
  }
\fi

\unless\ifjournal
\maketitle
\fi

\begin{abstract}
This paper provides an overview of the 2.0 release of the ArborX library, a performance portable geometric search library based on Kokkos. We describe the major changes in ArborX 2.0 including a new interface for the library to support a wider range of user problems, new search data structures (brute force, distributed), support for user functions to be executed on the results (callbacks), and an expanded set of the supported algorithms (ray tracing, clustering).
\end{abstract}

\ifjournal
\maketitle
\fi

\blfootnote {%
This manuscript has been authored by UT-Battelle, LLC, under contract
DE-AC05-00OR22725 with the U.S. Department of Energy. The United States
Government retains and the publisher, by accepting the article for
publication, acknowledges that the United States Government retains a
nonexclusive, paid-up, irrevocable, world-wide license to publish or reproduce
the published form of this manuscript, or allow others to do so, for United
States Government purposes. The DOE will provide public access to these results
in accordance with the DOE Public Access Plan
(http://energy.gov/downloads/doe-public-access-plan).
}

\section{Overview}\label{s:overview}

\arborx is a performance portable geometric search library. \arborx version 2.0 was released April 16, 2025. This paper provides an overview of the new features developed in \arborx since the original paper~\cite{arborx2020} published in 2020, and serves as a citable reference for the \arborx software library version 2.0. It is available for free under the terms of \emph{BSD 3-Clause} license. It is available on GitHub at \url{https://github.com/arborx/ArborX}.

The major changes in \arborx since~\cite{arborx2020} are: 
\begin{itemize}
  \item New interface for search indexes (\Cref{s:interface})
  \item Support for user functions called on a match (callbacks) (\Cref{s:callbacks})
  \item Support for distributed search using MPI (\Cref{s:distributed})
  \item New clustering algorithms (\Cref{s:clustering})
  \item Support for ray-tracing (\Cref{s:ray_tracing})
  \item Multiple performance improvements (\Cref{s:performance})
\end{itemize}

While the major changes are discussed in detail in \Cref{s:major_changes},
there are several other noteworthy changes included in the 2.0 release, which we briefly outline here:
\begin{itemize}
  \item Modernization of the code and build system.

    The minimum requirements for \arborx are now C++20, CMake 3.22 and Kokkos 4.5.
  \item New brute-force search structure
  \item Wider dimensionality and precision support for geometries.

    Geometries now support different data dimensions (1-10) and floating point precisions.
  \item Increased support for AMD and Intel GPUs (through the developments in Kokkos HIP and SYCL backends).
  \item Expanded list of supported geometries.

    In addition to previously supported points, axis-aligned bounding boxes and spheres, \arborx now supports $k$-DOPs \citep{klosowski1998}, triangles, rays, tetrahedrons, and segments.
  \item New implementation of the moving least squares algorithm \citep{quaranta2005} as part of the interpolation subpackage.
\end{itemize}

\section{Major changes to the library}\label{s:major_changes}

This release contains a number of significant changes, which we will discuss in
this Section. A number of these changes required breaking backwards compatibility.

\subsection{Updates to interface}\label{s:interface}
\subsubsection{Original interface (API v1)}

The main search data structure (index) in \arborx is \code{BVH} (Bounding Volume Hierarchy). Its original interface looked like the
following:
\begin{minted}{cpp}
template <typename MemorySpace>
class BVH {
public:
 using memory_space = MemorySpace;
 using size_type = typename MemorySpace::size_type;
 using bounding_volume_type = ArborX::Box;

 template <typename Primitives>
 BVH(Primitives const &primitives);

 template <typename Predicates>
 void query(Predicates const &predicates, Kokkos::View<int *, MemorySpace> &indices,
            Kokkos::View<int *, MemorySpace> &offset) const;

 KOKKOS_FUNCTION size_type size() const;
 KOKKOS_FUNCTION bool empty() const;
 KOKKOS_FUNCTION bounding_volume_type bounds() const;
};
\end{minted}

The following example shows how the interface was used in practice:
\begin{minted}[fontsize=\footnotesize]{cpp}
// Create the View for the bounding boxes
Kokkos::View<ArborX::Box*, MemorySpace> boxes("boxes", num_boxes);
// Fill in the bounding boxes
...
// Create the bounding volume hierarchy
ArborX::BVH<MemorySpace> bvh(boxes);
// Create the View for the spatial queries
Kokkos::View<ArborX::Within *, MemorySpace> queries("queries", num_queries);
// Fill in the queries
...
// Perform the search
Kokkos::View<int*, MemorySpace> offsets("offsets", 0);
Kokkos::View<int*, MemorySpace> indices("indices", 0);
bvh.query(queries, indices, offsets);
\end{minted}
First, a user constructs bounding boxes for the data, and
calls the constructor. Next, the queries are built. Each query corresponds to a
pair of a query point and a number of neighbors to be found (nearest query), or
a pair of a query point and a radius (spatial query). Then, two views are
allocated to store the results, \code{indices} (the indices of the bounding
boxes that satisfy the queries) and \code{offsets} (the offsets
in~\code{indices} for each query)\footnote{This format is similar to that of
compressed sparse row format that is commonly used to store sparse matrices.}.
Finally, the search is done by invoking \code{BVH}'s  \code{query} function.

\subsubsection{Limitations of the original interface}

The original interface was designed to be easy to use while providing some limited flexibility.
With a growth in the number of users, and proliferation of desired search tasks for
different applications, it became important to reduce the existing limitations, which we 
briefly discuss below.

\paragraph{Hardcoded dimensionality.}
\arborx only supported 3D objects. It also supported 2D objects by treating them
as 3D through setting $z = 0$, at the cost of extra memory storage and slower
traversal.

\paragraph{Lack of execution spaces.}
The interface did not allow users to pass execution spaces as an argument.
Instead, it used the default Kokkos execution space. This made it impossible to
run multiple searches simultaneously, or running the search in parallel with
other user kernels. This also made any fencing global for a device, not a
single stream.

\paragraph{Lack of fine nearest neighbor search.}
Users were able to only expose the axis-aligned boxes to \arborx through the
interface. When searching for the nearest objects, ArborX only considered
distances to those boxes, and not to the user data. This meant that for user
geometries other than points and boxes, the results produced by \arborx corresponded to a coarse search.

\paragraph{Hardcoded bounding volume type.}
\arborx only supported axis-aligned bounding boxes (AABBs) as bounding volume.
There are many other bounding volumes used in practice and more efficient for
certain problems, such as oriented bounding boxes (OBBs), k-discrete oriented
polytopes (kDOPs)~\cite{klosowski1998}, and others. The choice of a bounding
volume is particularly important for distributed search queries, when this
directly impacts the amount of network communication.

\paragraph{No support for user functions (callbacks)}
Users are often interested in performing some operation on the results of each
query. For example, they may be only interested in the number of neighbors,
average distance, or updating some quantity. The results produced as a pair of
(\code{offsets}, \code{indices}) serve only as an intermediate step for that,
at the cost of extra memory reads and writes. Moreover, the number of the
results may exceed the memory capacity for particularly dense problems. In the distributed search using MPI, it additionally cuts down on the number of communication rounds and the amount of transferred data between different ranks. The
original interface did not allow users to specify an operation to execute on
the results without storing them.


\subsubsection{New interface (API v2)}

To address the mentioned limitations, the interface of \arborx was reworked
in version 2.0. The new interface of the search indexes was strongly inspired by Boost.Geometry.Index~\cite{boost_geometry}. An index may now store
user-provided objects of any type \code{Value}, as long as \arborx knows how to
extract a geometric object of type \emph{Indexable}. This extraction is done with a
helper functor \code{IndexableGetter} that can be specified by a user. This
allows \arborx to behave like a container while also providing a search index
over the provided data.

The current interface looks as follows:
\begin{minted}{cpp}
template <typename MemorySpace, typename Value, typename IndexableGetter, typename BoundingVolume>
class BVH {
 public:
  using memory_space = MemorySpace;
  using size_type = typename MemorySpace::size_type;
  using bounding_volume_type = BoundingVolume;
  using value_type = Value;

  template <typename ExecutionSpace, typename Values>
  BVH(ExecutionSpace const &space, Values const &values,
      IndexableGetter const &indexable_getter = IndexableGetter());

  template <typename ExecutionSpace, typename Predicates, typename Callback>
  void query(ExecutionSpace const &space, Predicates const &predicates,
             Callback const &callback); // (1)
  template <typename ExecutionSpace, typename Predicates, typename Callback>
  void query(ExecutionSpace const &space, Predicates const &predicates,
             Callback const &callback,
             Kokkos::View<ValueLike *, MemorySpace> &values,
             Kokkos::View<int *, MemorySpace> &offset) const; // (2)
  template <typename ExecutionSpace, typename Predicates>
  void query(ExecutionSpace const &space, Predicates const &predicates,
             Kokkos::View<Value *, MemorySpace> &values,
             Kokkos::View<int *, MemorySpace> &offset) const; // (3)

  KOKKOS_FUNCTION size_type size() const noexcept;
  KOKKOS_FUNCTION bool empty() const noexcept;
  bounding_volume_type bounds() const noexcept;
};
\end{minted}
The major differences with the original interface are:
\begin{itemize}
  \item Updated template arguments. A search index is now templated on a Kokkos memory space, stored value type, indexable getter and bounding volume.
  \item Taking in a Kokkos execution space for both constructor and queries, allowing multiple searches to be performed in parallel.
  \item Introduction of an indexable getter argument in the constructor.
  \item Three different \code{query} calls:
    \begin{enumerate}
      \item[(1)] pure callback query that executes an operation on every match but does not store the results;
      \item[(2)] a callback query that does an operation on a match and stores the result of that operation. As this operation may change the type, the type of the results stored in the  \code{values} array may be different from \code{Value} type.
      \item[(3)] a simple storage query that is similar to APIv1
    \end{enumerate}
\end{itemize}
The \code{Values} in the constructor can be provided using the \code{ArborX::AccessTraits} mechanism as before.
Together with callbacks (described in \Cref{s:callbacks}), this provides a
flexible interface, supporting many more of the use cases.

For example, the following code shows how to use the new interface:
\begin{minted}[fontsize=\footnotesize]{cpp}
// Create the View for the bounding boxes
Kokkos::View<ArborX::Box<3> *, MemorySpace>
      boxes("boxes", num_boxes);
// Fill in the bounding boxes
...
// Create an execution space
Kokkos::DefaultExecutionSpace space;
// Create the bounding volume hierarchy
ArborX::BVH const bvh(space, boxes);
// Create the View for the spatial queries
Kokkos::View<decltype(ArborX::intersects(ArborX::Point<3>{})) *, MemorySpace>
      queries("queries", num_queries);
// Fill in the queries
...
// Perform the search
Kokkos::View<int*, MemorySpace> offsets("offsets", 0);
Kokkos::View<ArborX::Box<3> *, MemorySpace> values("values", 0);
bvh.query(space, queries, values, offsets);
\end{minted}
As before, a user starts by constructing bounding boxes for the data. Note that the bounding boxes are now templated on the spatial dimension. Next, \code{BVH}'s constructor is called. Compared to the old API, the user needs to provide a Kokkos execution space that encapsulates the execution resources to use to perform the work. Next, the queries are built. Then, two views are allocated to store the results, \code{values} (the bounding boxes that satisfy the queries) and \code{offsets} (the offsets in \code{values} for each query). Finally, the search is performed by invoking the \code{BVH}'s \code{query} function. The main difference with the old API is that \code{query} returns the objects used to build the \code{BVH} object instead of their indices.

%

\subsection{Callbacks}\label{s:callbacks}
Callbacks are functors that are called on \code{values} that satisfy a predicate. Two different callbacks are supported. The first one does not produce any output (\emph{pure} callback):

\begin{minted}[fontsize=\footnotesize]{cpp}
struct Callback {
 template<typename Predicate, typename Value>
 KOKKOS_FUNCTION
 RT operator()(Predicate const &predicate, Value const &value) const;
};
\end{minted}

Here, \code{RT} is the return type, which can either be \code{void}, or a special type indicating early traversal termination.

The second callback type allows to produce output:

\begin{minted}[fontsize=\footnotesize]{cpp}
struct CallbackWithOutput {
 template<typename Predicate, typename Value, typename OutputFunctor>
 KOKKOS_FUNCTION
 void operator()(Predicate const &predicate, Value const &value,
                 OutputFunctor const &output) const;
};
\end{minted}
Users may use this callback combined with a general query (2) to finely control the stored output of each query. A user may compute and store any number of values for each positive match. These values may differ from the data stored in the index. For example, a user may desire to find all mesh elements that contain a given degree of freedom, interpolate a field from the found element, and store the interpolated value as an output.

The following example counts the number of results for each query up to a specified limit $N$. Once the number of results for a given query reaches the limit, the traversal for that query is terminated.
\begin{minted}[fontsize=\footnotesize]{cpp}
template<typename Counts>
struct CountUpToNCallback {
  Counts _counts;
  int _n;
  template<typename Predicate, typename Value>
  auto operator()(Predicate const &predicate, Value const &) const {
    int const i = ArborX::getData(query);
    if (_counts(i)++ >= _n)
      return ArborX::CallbackTreeTraversalControl::early_exit;
    return ArborX::CallbackTreeTraversalControl::normal_continuation;
  }
};
// Create an execution space
Kokkos::DefaultExecutionSpace space;
// Construct a hierarchy and queries
ArborX::BVH const bvh(space, ...);
Kokkos::View<ArborX::Within<...> *, MemorySpace> queries(...);
// Perform the search
constexpr int N = 10;
Kokkos::View<int*, MemorySpace> counts("counts", queries.size());
bvh.query(space, ArborX::attach_indices(queries), CountUpToNCallback{counts, N});
\end{minted}
Here, we used the \arborx helper function \code{attach\_indices} to
attach indices to a query. In general, \arborx allows attaching arbitrary data to
each query. The attached data is then obtained inside the callback through
\code{getData()} call. We store the counts in a temporary view. As queries are
not shared among threads, we do not need atomic operations when increasing the
count number for each query. The callback returns the \code{early\_termination}
value once the threshold is reached. This value terminates the traversal for that query.

\subsection{Distributed search}\label{s:distributed}

\arborx now provides a distributed search index \code{ArborX::DistributedTree}, with an interface closely following \code{ArborX::BVH} only adding an \code{MPI\_Comm} constructor argument.

The distributed search supports both spatial and nearest queries. Each processor constructs a local search index based on the provided data, and then a top level index providing an overview of the local bounding boxes. During the search, each processor provides a set of local queries that are then communicated to the processors that have the necessary data. The result is returned to the processor originating a query.

Distributed search fully supports callbacks. Callbacks are executed on the processor owning the data that satisfies a query. This allows to reduce communication required to execute an operation by avoiding crossing the network multiple times to execute operations. One could consider, for example, a distributed finite element mesh and asking to interpolate a solution on that mesh into a set of query points. Using callbacks achieves this without a need to communicate mesh elements among MPI ranks.

Distributed search can also take advantage of the GPU-aware MPI, if present, through the use of \code{ARBORX\_ENABLE\_GPU\_AWARE\_MPI} configuration option. When enabled, the data stays resident to accelerator memory instead of being transferred to host for communications.

The following example show a procedure to find $N$ closest points in a distributed setting:
\begin{minted}[fontsize=\footnotesize]{cpp}
struct PairIndexRank {
  unsigned index;
  int rank;
};
struct CallbackWithRank
{
  int _rank;
  template <typename Predicate, typename Value, typename Output>
  KOKKOS_FUNCTION
  void operator()(Predicate const &, ArborX::PairValueIndex<Value> const &value,
                  Output const &out) const {
    out({.index = value.index, .rank = _rank});
  }
};

// Create an execution space
ExecutionSpace exec;
// Set up an MPI communicator
MPI_Comm comm = MPI_COMM_WORLD;
// Find local rank
int comm_rank;
MPI_Comm_rank(comm, &comm_rank);
// Construct local data on each MPI rank
Kokkos::View<ArborX::Point<3> *, MemorySpace> points("points", n);
...
// Build a distributed tree
ArborX::DistributedTree tree(comm, exec, ArborX::attach_indices(points));

// Find N closest points for each local point and store them together with
// the rank they belong to
constexpr int N = 3;
Kokkos::View<PairIndexRank *, MemorySpace> values("indices_with_ranks", 0);
Kokkos::View<int *, MemorySpace> offsets("offsets", 0);
tree.query(exec, ArborX::make_nearest(points_device, N),
           CallbackWithRank{comm_rank}, values, offsets);
\end{minted}
We start with constructing a set of local points and proceed with
building a distributed index. As we mentioned before, the only difference in
the distributed constructor's interface is the presence of \code{MPI\_Comm} as
the first argument. Internally, the provided communicator is duplicated so that
\arborx's internal communication is isolated from the user application. We attach
the indices to the data, and then perform the search. We use another helper
function here, \code{make\_nearest}, to easily construct a set of nearest
queries for the data. We also use a \code{CallbackWithRank} callback, as
by default \arborx only returns the data stored in the tree. The result is then
stored as pairs of an index with the rank it was found on.

\subsection{Clustering algorithms}\label{s:clustering}
\arborx now provides access to several clustering algorithms:

\begin{itemize}
    \item \emph{Density-Based Spatial Clustering of Applications with Noise (DBSCAN)}

    DBSCAN \citep{ester1996} is a density-based clustering algorithm. Compared to other techniques like $k$-means, it can deal with clusters of arbitrary shapes and does not require knowing the number of clusters \emph{a priori}.

    \arborx provides two implementations of the algorithm depending on the sparsity of the data (with respect to the DBSCAN's
    $\varepsilon$ parameter). A user may choose to use either FDBSCAN (for very sparse data) or FDBSCAN-DenseBox (for data with dense regions) algorithms. Both algorithms are implemented to run efficiently on both CPUs and GPUs, and have been used in production for large scale cosmology problems~\citep{arborx_dbscan_2025}. For more information, see~\cite{arborx_dbscan_2023}.

    The work to implement DBSCAN algorithm in the distributed setting is ongoing.

    \item \emph{Euclidean minimum spanning tree (EMST)}

    EMST computes the minimum spanning tree of the distance graph of a set of points, i.e., a graph where each pair of vertices are connected by an edge of weight equal to the distance between them. EMST also serves as a basis for the HDBSCAN* algorithm \citep{campello_2015}. \arborx provides a new implementation of the algorithm suitable for GPUs. For more information, see \cite{arborx_emst}.
\end{itemize}

As a simple example, consider running DBSCAN algorithm for a given set of 2D points:
\begin{minted}[fontsize=\footnotesize]{cpp}
// Create an execution space
ExecutionSpace exec;
// Construct points
Kokkos::View<ArborX::Point<2> *, MemorySpace> points("points", n);
...
// Run DBSCAN
constexpr float eps = 1.5;
constexpr int minPts = 5;
auto labels = ArborX::dbscan(exec, points, eps, minPts);
\end{minted}
The result is stored in a 1D integer array \code{labels}. If a given
label is equal to -1, the corresponding point is a noise point. Otherwise, the
label corresponds to a cluster index.

\subsection{Ray tracing support}\label{s:ray_tracing}
\arborx now supports ray tracing. While there are multiple ways to define a ray, an \code{ArborX::Ray} is defined by its origin and its direction. Ray tracing is currently supported by a limited set of geometries: a box, a triangle and a sphere. \arborx provides three kinds of predicates for ray tracing:
\begin{itemize}
  \item \code{nearest}: this predicate finds the first $k$ geometric objects that the rays intersect. One could consider the rays to be absorbed after $k$ collisions. The case $k = 1$ corresponds to finding the closest geometric object that the ray intersects. 
  \item \code{intersect}: this predicate finds all geometric objects that the rays intersect. It is equivalent to the \code{nearest} predicate where $k$ is infinity. This can be likened to the rays intersecting perfectly transparent objects.
  \item \code{ordered\_intersect}: this predicate is similar to \code{intersect} but the intersected objects are ordered as if a physical ray would encounter them. This can be useful when a ray is depositing energy through a medium and it is necessary to keep track of the history of the intersections.
\end{itemize}
The \code{nearest} and \code{intersect} predicates are also supported in the case of distributed ray tracing where the geometric objects and the rays exist on different MPI ranks.

\subsection{Performance improvements}\label{s:performance}
\arborx's internal implementation of the bounding volume hierarchy (BVH) was modified significantly in this release, dramatically boosting its performance. We highlight the following changes among many others:
\begin{itemize}
    \item Morton codes used during the construction changed from 32-bit to 64-bit by default.
    \item The hierarchy construction algorithm switched from Karras' \citep{karras2012} to Apetrei's \citep{apetrei2014} algorithm.
    \item Spatial search is now performed using stackless algorithm \cite{arborx_apetrei}.
    \item Search for pairs of objects now uses a special traversal algorithm (see \cite{arborx_dbscan_2025}).
    \item Search can now be terminated early through the use of callbacks (see \Cref{s:callbacks}).
    \item Sorting in both construction and search now takes advantage of the vendor-specific libraries (\texttt{Thrust} for CUDA, \texttt{rocThrust} for HIP, and \texttt{oneDPL} for SYCL).
\end{itemize}

In \Cref{t:perf}, we compare the performance between two versions of \arborx:
the ``old'' version (hash 02f9a81, March 2020) and the ``new'' version (version
2.0, hash 041bf18e, April 2025), compiled against Kokkos 4.7\footnote{The old
version required minor patching to work with a new version of Kokkos.}, using
CUDA 12.4. For our experiments, we used a cosmology dataset
from~\cite{arborx_dbscan_2025} (Section 4) containing $\approx37M$ points. We
ran the construction and search procedure (with radius $r = 0.042$) on a Nvidia
A100 GPU. As we can see, using Thrust instead of the native \code{Kokkos::BinSort}
algorithm is critical. Further $5\times$ improvements are coming from the
stackless traversal algorithm as well as from storing less data (leaf nodes
containing points now take 16 bytes instead of 32). When comparing old and new
versions, all found results were stored. We also display the performance of the query
when the results are simply counted using a pure callback. It shows an
additional improvement of over $2\times$.
\begin{table}[t]
  \caption{Performace comparison of old and new \arborx versions on a $\approx37M$ cosmology
  dataset on Nvidia A100 GPU.\label{t:perf}}
  \centering
  \begin{tabular}{lccc}
    & Old (no Thrust) & Old (with Thrust) & New \\
  \toprule
    Construction  & 17.006 & 0.021 & 0.018 \\
    Query         & 18.619 & 1.272 & 0.257 \\
    Query (count) &      - &     - & 0.124 \\
  \bottomrule
  \end{tabular}
\end{table}

\ifjournal
\begin{acks}
\else
  \section*{Acknowledgements}
\fi
Other than the authors of this paper, the following people contributed code to
this release: Yohann Bosqued, Jared Crean, Ana Gainaru, Wenjun Ge, Christoph Junghans, Shihab Shahriar Khan, Phil Miller, Tuan Pham.
Their contributions are much appreciated!

This research was supported by the Exascale Computing Project (17-SC-20-SC), a
collaborative effort of the U.S. Department of Energy Office of Science and
the National Nuclear Security Administration.
\ifjournal
\end{acks}
\fi

\ifjournal
  \bibliographystyle{ACM-Reference-Format}
\else
  \bibliographystyle{apalike}
\fi

\bibliography{main}

\end{document}